\documentclass[iop]{emulateapj}
\usepackage{graphicx}
\usepackage{amssymb}

\shorttitle{WFC3 spectroscopy of Cl J1449+0856}
\shortauthors{R. Gobat et al.}

\begin{document}

\title{WFC3 grism confirmation of the distant cluster Cl J1449+0856 at $\langle \lowercase{z}\rangle=2.00$:\\
Quiescent and star-forming galaxy populations}

\author{R. Gobat\altaffilmark{1},V. Strazzullo\altaffilmark{1},E. Daddi\altaffilmark{1},
M. Onodera\altaffilmark{2},M. Carollo\altaffilmark{2},A. Renzini\altaffilmark{3},
A. Finoguenov\altaffilmark{4},A. Cimatti\altaffilmark{5},C. Scarlata\altaffilmark{6},N. Arimoto\altaffilmark{7}
}

\altaffiltext{1}{Laboratoire AIM-Paris-Saclay, CEA/DSM-CNRS--Universit\'e Paris Diderot, 
Irfu/Service d'Astrophysique, CEA Saclay,Orme des Merisiers, F-91191 Gif sur Yvette, France}
\altaffiltext{2}{Institute for Astronomy, ETH Z\"{u}rich, Wolfgang-Pauli-strasse 27, 8093 Z\"{u}rich, 
Switzerland}
\altaffiltext{3}{INAF - Osservatorio Astronomico di Padova, Vicolo dell'Osservatorio 5, I-35122 Padova, 
Italy}
\altaffiltext{4}{University of Helsinki, P.O. Box 33 (Yliopistonkatu 4), FI-00014 Helsinki, Finland}
\altaffiltext{5}{Universit\`{a} di Bologna, Dipartimento di Astronomia, Via Ranzani 1, I-40127 Bologna, 
Italy}
\altaffiltext{6}{School of Physics and Astronomy, University of Minnesota, 116 Church Street Southeast, 
Minneapolis, MN 55455, USA}
\altaffiltext{7}{Subaru Telescope, National Astronomical Observatory of Japan, 650 North A'ohoku Place, Hilo, 
HI 96720, USA}

\begin{abstract}
We present deep \emph{Hubble Space Telescope} Wide Field Camera 3 (\emph{HST}/WFC3) slitless spectroscopic observations of the 
distant cluster Cl J1449+0856. These cover a single pointing with 18 orbits of G141 spectroscopy and F140W imaging, allowing us 
to derive secure redshifts down to $M_{140}\sim25.5$~AB and 3$\sigma$ line fluxes of $\sim5\times10^{-18}$~erg s$^{-1}$ cm$^{-2}$. 
In particular, we were able to spectroscopically confirm 12 early-type galaxies (ETGs) in the field up to $z\sim3$, 6 
of which in the cluster core, which represents the first direct spectroscopic confirmation of quiescent galaxies in a $z=2$ 
cluster environment. With 140 redshifts in a $\sim6$~arcmin$^2$ field, we can trace the spatial and redshift galaxy distribution 
in the cluster core and background field. We find two strong peaks at $z=2.00$ and $z=2.07$, where only one was seen in our 
previously published ground-based data. Due to the spectroscopic confirmation of the cluster ETGs, we can now 
re-evaluate the redshift of Cl J1449+0856 at $z=2.00$, rather than $z=2.07$, with the background overdensity being revealed 
to be sparse and ``sheet''-like. This presents an interesting case of chance alignment of two close yet unrelated structures, 
each one preferentially selected by different observing strategies.
With 6 quiescent or early-type spectroscopic members and 20 star-forming ones, Cl J1449+0856 is now reliably confirmed to be 
at $z=2.00$. The identified members can now allow for a detailed study of galaxy properties in the densest environment at $z=2$.
\end{abstract}

\keywords{galaxies: clusters (Cl J1449+0856)---galaxies: high-redshift}

\section{\label{intro}Introduction}

Galaxy clusters are the most overdense structures in the Universe and as such invaluable tools to constrain cosmological parameters 
and to understand how the local environment can bias galaxy evolution, in particular the 
formation and evolution of massive early-type galaxies (ETGs). This galaxy population dominates cluster cores at $z<1.5$ and, 
while its formation and assembly over cosmic time is still not yet fully understood, it has nevertheless been found to have a 
very old component, with a substantial part of its stellar mass having formed at redshifts in excess of two 
\citep[e.g.,][]{Thom05,Mei09,Ret10}. The question of its origin, and relation with the formation of the cluster environment 
itself, can thus be more easily addressed at high redshift, closer to the formation of the first ETGs and into the epoch of 
protoclusters, when the first relaxed galaxy clusters start appearing and the beginnings of the red sequence become visible 
\citep{Kod07,Zir08}.\\
Distant clusters have traditionally been discovered using the characteristic signature of their hot gas atmosphere, either 
directly through its X-ray emission \citep[e.g.,][]{Ros98,Rom01} or its effect on the cosmic microwave background \citep{SZ,Bar96}. 
This type of selection however requires massive, relaxed structures and, in the case of X-ray surveys, is limited 
by surface brightness. 
Over the past 15 years, many galaxy clusters have been found at $z<1.5$ using these techniques 
\citep[e.g.,][]{Stan97,Ros04,Mul05,Stan06,Tan10,Fas11,San11,Rei13}. However, instrumental limitations and the sharply decreasing 
number density at high redshift of the type of massive structure detectable by X-ray or Sunyaev Zel'dovich observations conspire 
to progressively reduce the number of galaxy clusters that can be reliably identified at $z>1.5$. On the other hand, the photometric 
selection of overdensities of massive galaxies at high redshift \citep[e.g.,][]{Gla00,Eis08,Wi08} has been quite successful in 
discovering clusters above $z=1.5$ \citep[e.g.,][]{And09,Pap10,Gob11,Stan12,Spi12,Muz13}. 
Yet, most of such high-redshift structures are of relatively low mass, their number, despite the recent progress, is still limited 
and they require spectroscopic confirmation. While it is possible in exceptional cases to determine a cluster's redshift directly 
from X-ray observations \citep[e.g.,][]{Toz13}, typical cluster identifications usually need to be followed up by spectroscopic 
observations, lest their usefulness as cosmological probes or galaxy evolution laboratories be limited. 
Spectroscopic confirmation of high-redshift cluster members, and especially quiescent galaxies, is a difficult task, but required 
to convincingly demonstrate that a structure is genuinely evolved, with an established population of old galaxies. 
In this case, spectroscopic confirmation implies detection of the stellar continuum of the central ETG population and thus 
requires a high enough signal-to-noise ratio (S/N), which is only achievable from the ground with very long integration times, either 
in the optical \citep[e.g.,][]{Cim08} or in the near-infrared \citep[NIR; e.g.,][]{Ono12}. Further complicating matters, the main spectral 
feature of old stellar populations, the 4000\AA~break, is at $z>1.5$ redshifted into the NIR, requiring a sensitivity in this 
wavelength range that, until recently, only few instruments could provide.\\
At $z>2$, on the other hand, efforts have mostly been focused on finding galaxy cluster progenitors, often around giant radiogalaxies, 
as overdensities of star-forming objects selected by their line emission \citep[e.g.,][]{Pen97,Ste05,Ov06,Ven07,Ha11,Ko13}. 
These protoclusters, being readily identified in narrow-band surveys, are thus paradoxically easier to confirm than the later (and 
closer) galaxy clusters. This biases our view of the $z\gtrsim2$ universe, as we then miss the more quiescent and evolved structures 
\citep[e.g.,][]{Spi12} that may be present at this epoch.\\

In contrast with ground-based instruments, the \emph{Hubble Space Telescope} Wide Field Camera 3 (hereafter \emph{HST}/WFC3) in 
its grism mode does not suffer from atmospheric absorption and OH glow. Its great sensitivity, coupled with a relatively low spectral 
resolution, allows for efficient continuum detection. Its capacity to observe, in principle, every object in its field of view also 
allows for relatively unbiased surveys \citep[e.g.,][]{Tru11}. Thus, even with its relatively limited field of view 
($\sim2$\arcmin$.3\times2$\arcmin), the \emph{HST}/WFC3 is an ideal instrument for obtaining spectra of compact, passively 
evolving galaxies in cluster cores, in redshift ranges traditionally inaccessible to ground-based instruments \citep[e.g.,][]{Stan12}.\\

In this work we present deep \emph{HST}/WFC3 observations of the infrared-selected galaxy cluster Cl J1449+0856 \citep{Gob11}, 
the most distant confirmed so far and also the most distant with a detected X-ray emission. This field benefits from a large 
multiwavelength dataset which has since been significantly expanded, both at short (UV) and longer (far-IR to 
radio) wavelengths. Here we concentrate on the \emph{HST}/WFC3 spectroscopic data, their analysis and the redshifts obtained from 
them. A more detailed study of other galaxy properties derived from the spectra will be presented in forthcoming papers. 
In particular, the UV to NIR parts of this dataset have recently been reanalyzed with the inclusion of the high-resolution 
imaging and spectroscopic redshifts presented here. All this has been used to derive new accurate photometric redshifts and 
make a comparative study of the star-forming and quenched galaxy populations of the cluster. These results are presented in a 
companion paper \citep{Stra13}.\\
The field around Cl J1449+0856 had already been spectroscopically observed from the ground with the FORS2 and VIMOS instruments 
on the Very Large Telescope \citep[VLT;][]{Gob11} and MOIRCS on Subaru, targeting mostly sBzK-selected \citep{Dad04} galaxies. 
However, redshifts could be measured only for star-forming galaxies and none of the objects spectroscopically confirmed by these 
ground-based observations was found in the cluster core. For the red galaxies dominating the core, only photometric redshifts were 
available. 
As the distribution of spectroscopic redshifts in the surrounding area was strongly peaked at $z=2.07$, and that of the photometric 
redshifts of red galaxies also peaked at $z_{phot}\simeq2.05$, the redshift of $z=2.07$ was attributed to the cluster. Thus, despite 
the convincing nature of Cl J1449+0856, a full and definitive confirmation of the redshift of the core galaxy population could not be 
obtained.To remedy to this situation, we carried out deep \emph{HST} spectroscopy of the central galaxy overdensity with the red G141 
grism. This allowed us to spectroscopically confirm both quiescent and star-forming galaxies in and around the cluster core, which led 
to a slight but significant revision of the cluster redshift.\\

Throughout the paper, we assume a $\Lambda$CDM cosmology with 
$H=70$~km~s$^{-1}$~Mpc$^{-1}$, $\Omega_m=0.27$ and $\Lambda=0.73$. Magnitudes are given in the AB photometric system throughout.

\begin{figure*}
\centering
\includegraphics[width=0.8\textwidth]{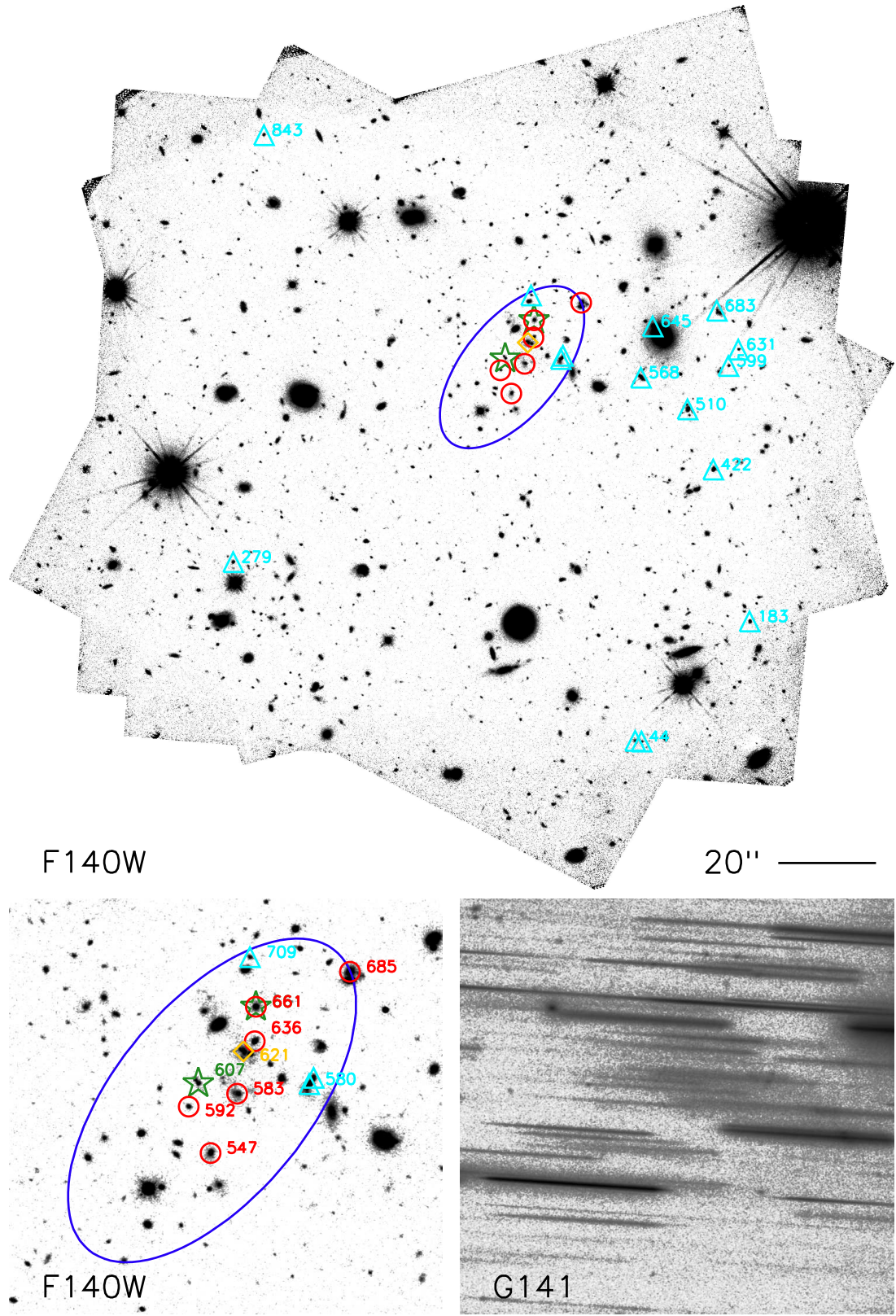}
\caption{Top: \emph{HST}/WFC3 image of Cl J1449+0856 in the F140W filter showing the pointing strategy, with the core 
circumscribed by a blue ellipse. The red circles, light blue triangles and orange diamonds mark the position of spectroscopic 
members, respectively quiescent, star-forming line-emitting and dusty star-forming.
The green stars indicate AGNs detected in the soft (bottom) and hard (top, tentative detection) \emph{Chandra} bands. 
Bottom left: \emph{HST}/WFC3 F140W cutout of the central region of the cluster. Bottom right: G141 grism image of 
the same region showing the slitless spectra. The grism frame was shifted so that the positions of the sources correspond to the 
center of their spectral traces.}
\label{fig:grism}
\end{figure*}

\begin{deluxetable}{ccccccc}
\centering
\tablecaption{WFC3 Observations\label{tab:obs}}
\tablehead{\colhead{Date} & \colhead{Angle} & \colhead{Orbits} & \colhead{Time} & \colhead {Time}\\
\colhead{} & \colhead{} &  \colhead{} & \colhead{(direct)} & \colhead{(grism)}\\
\colhead{} & \colhead{(deg)} &  \colhead{} & \colhead{(hr)} & \colhead{(hr)}}
\startdata
2010 26 Jun, 1 Jul&-4.5&10&0.6&7\\
2010 9 Jul&-28&4&0.3&2.7\\
2010 6 Jun&14.8&4&0.3&2.7
\enddata
\end{deluxetable}

\section{\label{obs}\emph{HST}/WFC3 observations}

The distant cluster Cl J1449+0856 was observed with the \emph{HST}/WFC3 using the G141 grism and F140W filter between 2010 June 
and July (see Table \ref{tab:obs}). The direct imaging was taken, as customary, to provide information on source positions and 
morphologies, to allow for the modeling of the spectra and to facilitate their extraction. 
Because of the high density of sources in this field, three different orientations were chosen and the dithering of the F140W imaging 
was set so that its total coverage be larger than any of the three grism pointings. This ensured that spectral contamination could be 
estimated with sub-pixel precision, also for traces originating outside the field of view of the individual pointings. 
At 18 orbits, this G141 data are some of the deepest yet. They cover a total area of 6.4~arcmin$^2$, with uniform coverage on 
$\sim3$~arcmin$^2$ (and $\sim4$~arcmin$^2$ covered by at least three orientations). 
Figure \ref{fig:grism} shows the Cl J1449+0856 field in direct and grism imaging, along with the positions of cluster 
members. Here we define the cluster core as a $\sim40$\arcsec$\times20$\arcsec region including the centroid of the X-ray emission and 
corresponding to the overdensity of red galaxies as initially found in \citet{Gob11} and confirmed in \citet{Stra13}. It also contains 
all the early-type members which we could spectroscopically confirm.\\
The data were in a first time reduced using the best available calibration files at the time and the standard aXe pipeline \citep{Kum09}. 
The F140W frames were combined with Multidrizzle and the resulting image was used to extract photometry. The multiband catalog, 
including ground-based observations, is described in \citet{Stra13}. The grism spectra in each orientation were reduced using 
the aXe software to produce cutouts of two-dimensional (2D) spectra for each source in the photometric catalog. 
As described in \citet{Gob12}, the individual grism frames were sky-subtracted separately. This was deemed necessary, 
as the density of traces in the grism images resulted in oversubtraction by the pipeline. As the version of aXe used here 
(v2.1) did not allow for correcting for cosmic rays independently of sky subtraction, these and residual bad pixels were masked a 
posteriori in each spectrum during the analysis (even tough, due to the low and variable resolution of the spectra, the aspect of 
such artifacts is quite different than that of actual spectral features). 
Finally, extracted one-dimensional (1D) spectra were flux-calibrated using the G141 first-order sensitivity function, rebinned for 
each spectrum to its pixel scale and smoothed to its effective resolution. Wavelength calibration, on the other hand, was handled by 
aXe. The limit for extraction was set at $m_{140}=25.5$, close to the $10\sigma$ limit \citep[1'' aperture,][]{Stra13} of the direct 
imaging and corresponding to a spectral continuum S/N of $\sim3$ per resolution element. This ensured that the initial sample be 
complete and allowed for easier component separation in the case of blended spectra (sources with S/N$<3$ were still included in 
the contamination estimates, but not extracted; see below).\\

\subsection{\label{ext}Extraction of Spectra}

Because the \emph{HST}/WFC3 instrument operates in slitless mode, two effects conspire to make the reduction and analysis of the 
spectra more challenging: while the G141 grism has a resolution of $R\simeq130$ (with a pixel size of 46.5\AA~and 0\arcsec.13~in 
the cross-dispersion, i.e., spatial, direction), each spectrum is effectively a convolution of an ``intrinsic'' 
spectrum with the spatial profile of the source in the dispersion (wavelength) direction. The actual spectral resolution 
can thus be significantly lower for extended sources. This results in broadened absorption or emission features and a lower 
S/N \citep[e.g.,][]{Bram12}, rendering their identification and interpretation sometimes difficult. More importantly, a spectrum can 
be contaminated by spectral traces originating from sources elsewhere in the field along the dispersion direction, including higher 
(first- and second-) or lower (zeroth) order modes usually cut off or well-separated by ground-based instruments. 
While the response of the grism in these higher orders is at most 30\% that of the first order in the usable wavelength range 
of the G141 grism ($1.1$~$\mu$m$<\lambda<1.7$~$\mu$m), and often an order of magnitude lower, higher-order contaminating traces 
from bright objects can still overwhelm the spectra of faint sources. Even when contamination is not dominant, it can 
negatively affect spectra by changing the overall shape of the continuum or adding spurious features (e.g., emission lines), resulting 
in misidentifications and erroneous redshifts. While observations are usually done in at least two different orientations to mitigate 
this problem, here the density of sources around Cl J1449+0856 and the depth of the data ensure that a significant fraction of 
spectra be contaminated in all three orientations.\\

There are several ways to correct for contamination, which involve spectral modeling of the contaminating 
traces\footnote{See, e.g., http://www.stsci.edu/hst/wfc3/analysis/grism\_obs/\allowbreak{}cookbook.html}.
At the time of this writing, the most sophisticated procedure implemented in aXe, the ``fluxcube'' method, uses imaging in bands bracketing 
the grism, interpolating them to model both the spectrum of sources and the spectral dependence of their spatial profiles. The fluxcube 
method can be somewhat costly to implement, however, as it requires \emph{HST} imaging in several filters with matching depths. As a 
compromise, trace profiles in the cross-dispersion direction can also be approximated as Gaussian using information yielded by 
the direct image. In the course of the reduction, we have tried both the Gaussian approximation of spectral profiles, which we found 
largely overestimated contamination, and the fluxcube method. In the latter case, as a workaround to the fact that \emph{HST} imaging was 
only available in one band, we used the F140W image to simulate \emph{HST} images at different wavelengths by re-scaling each object, using 
the ground-based photometry and the segmentation image produced by SExtractor \citep{SEx} as a mask.\\

\begin{figure}
\centering
\includegraphics[width=0.49\textwidth]{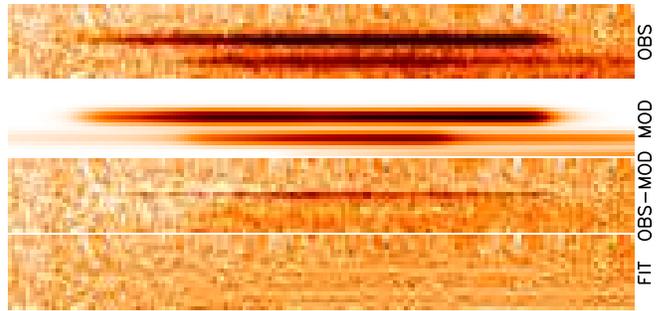}
\caption{From top to bottom: \emph{HST}/WFC3 two-dimensional spectrum of a quiescent galaxy, trace, and contamination model produced by 
aXe, residual spectrum after subtraction of the aXe model, and residuals after extraction by profile decomposition.}
\label{fig:2d}
\end{figure}

\begin{figure*}
\centering
\includegraphics[width=0.49\textwidth]{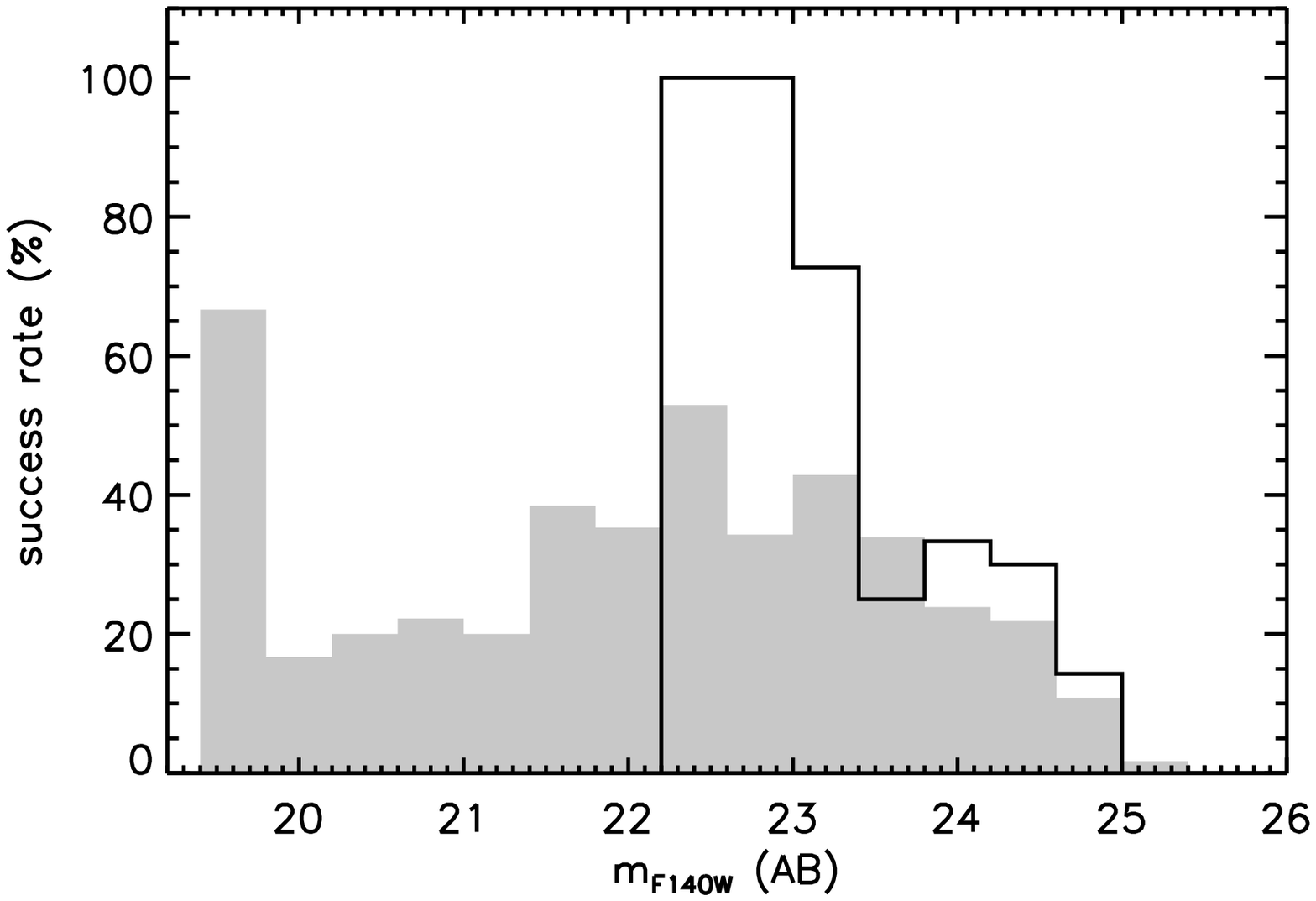}
\includegraphics[width=0.49\textwidth]{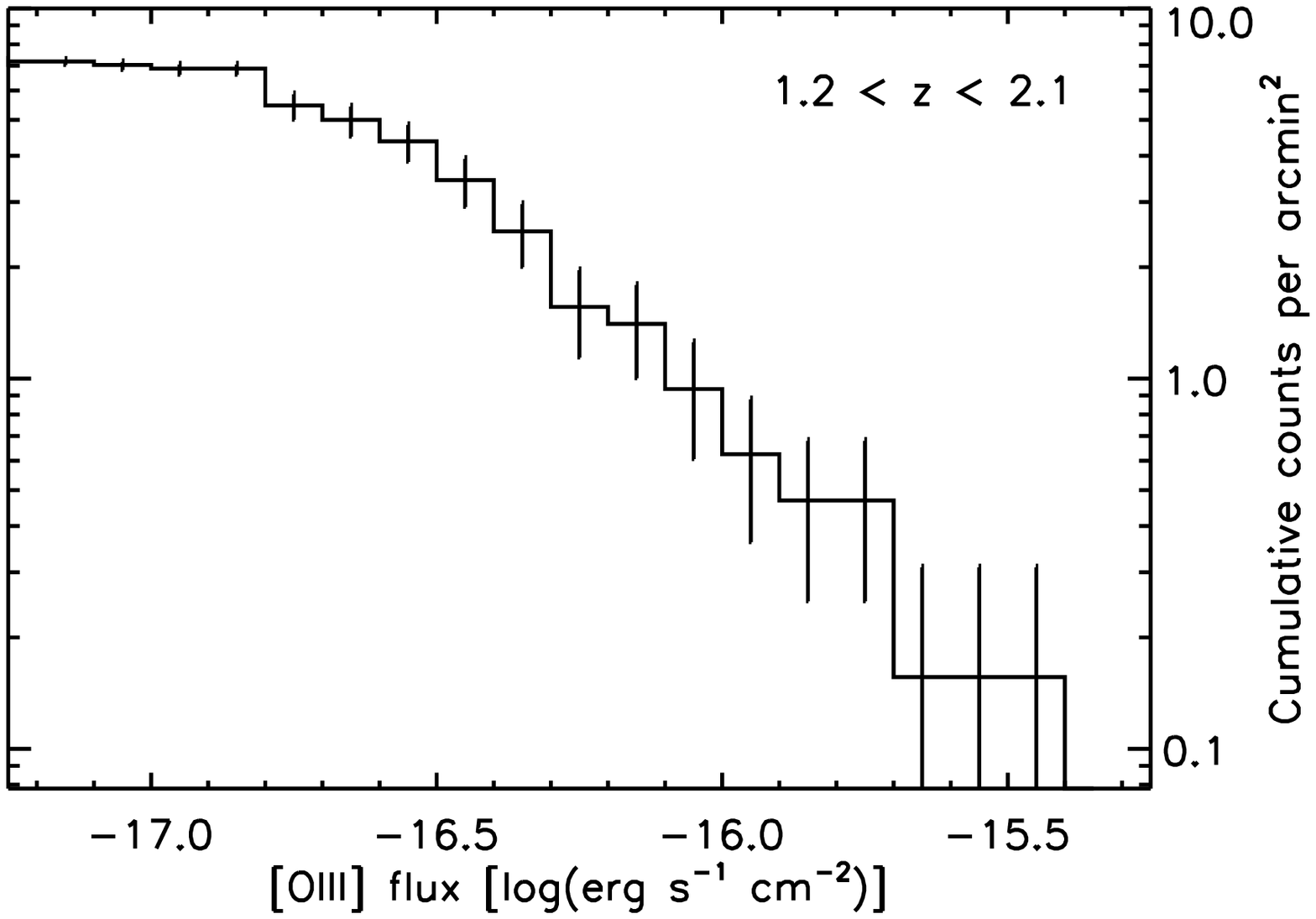}
\caption{Left: Fraction of redshifts compared to the total number of sources, as a function of F140W magnitude, 
in the full 6.4~arcmin$^2$ field (gray shaded histogram) and within the cluster core (black histogram) as delimited in 
Figure \ref{fig:2d}. Right: cumulative distribution of [OIII] line fluxes ($>3\sigma$ detections) between $z=1.2$ and $z=2.1$, 
excluding cluster members.}
\label{fig:success}
\end{figure*}

We found that this latter method worked well for sources with relatively high S/N. However, the contamination estimate is, in this 
case, entirely dependent on the quality of the photometry and can be biased by photometric uncertainties. Since it does not use true 
multiband \emph{HST} imaging, information on wavelength-dependent morphological variation is also lacking. Even when other \emph{HST} 
images are available, the quality of the spectral model will be set by the photometric sampling, especially around strong features like 
breaks which, because of the width of the filters, will be smoothed out. The contamination model is therefore always an (often poor) 
approximation of the real spectrum. However, calibration of the continuum is essential for those objects that have no or weak line 
emission, like quenched or highly reddened star-forming galaxies. The slitless spectra of these sources, especially faint ones, are thus 
at risk of being lost because of this sub-optimal contamination model. Where the type and distance of a contaminating source are known, 
such as in the case of a star, a more accurate template spectrum can be used in lieu of the photometric based model. However, this is 
almost never the case, the vast majority of sources in a grism observation lacking prior spectral and redshift information.\\

To address this problem, we also tried another approach which forwent the models produced by aXe, using only the science and error 
extensions of the 2D cutout, and dealt with contamination at the time of extraction. This was done by fitting each pixel column in 
the cross-dispersion direction with a combination of trace profiles derived from the direct images:\\
$\bullet$ For any given 2D spectrum, the positions of the traces (target and contaminants) falling in part or in full within the 
aperture were computed from their relative positions in the direct image.\\
$\bullet$ Object profiles in the dispersion direction were then estimated from the direct image, using the segmentation map to 
separate them. To account for dependence of the PSF with wavelength, these profiles were broadened (respectively sharpened) 
at longer (respectively shorter) wavelengths than the F140W pivot. The variation of the PSF across the wavelength range of the 
grism was measured from the uncontaminated traces of unsaturated stars in the field.\\
$\bullet$ The maximum continuum flux of the traces was estimated using the sources' spectral energy distributions 
\citep[SEDS; see][for a description of the photometric catalog]{Stra13} and the sensitivity curve for the corresponding spectral order. 
The background in the 2D spectrum was then estimated by masking the traces where they were expected to rise above the noise, 
allowing for a linear variation along the dispersion direction.\\
$\bullet$ Each pixel column in the cutout was then fit as a combination of profiles at their expected positions. The resulting 
spectrum is then given at each pixel by the normalization of each source profile and the error spectrum by the uncertainty on 
the fitted parameters.\\

For unblended spectra, the second approach yields results similar to the pipeline's untilted extraction although, by not making 
assumptions on spectral shapes, the residuals after subtraction of source and contamination spectra are lower than with trace 
models produced by aXe (see Figure \ref{fig:2d}). In the case of overlapping spectra, however, it can distinguish between features 
(emission lines, deep absorption lines or spectral breaks) originating from one source or the other based on the trace 
profiles, provided the relative brightness difference be not too large (on the other hand, one limitation of this approach is that 
here we assume a consistent source profile and thus, e.g.,, that emission features are distributed more or less evenly across the 
galaxies). We used this approach when the 2D spectra were severely contaminated or lacked emission lines (e.g., in the case of quiescent 
or heavily reddened systems). The extracted, contamination-subtracted 1D spectra in each orientation were first compared 
to each other, to check that they were consistent and then stacked before calibration. They were weighted according to exposure and 
wavelength-dependent contamination.\\

\begin{figure*}
\centering
\includegraphics[width=0.98\textwidth]{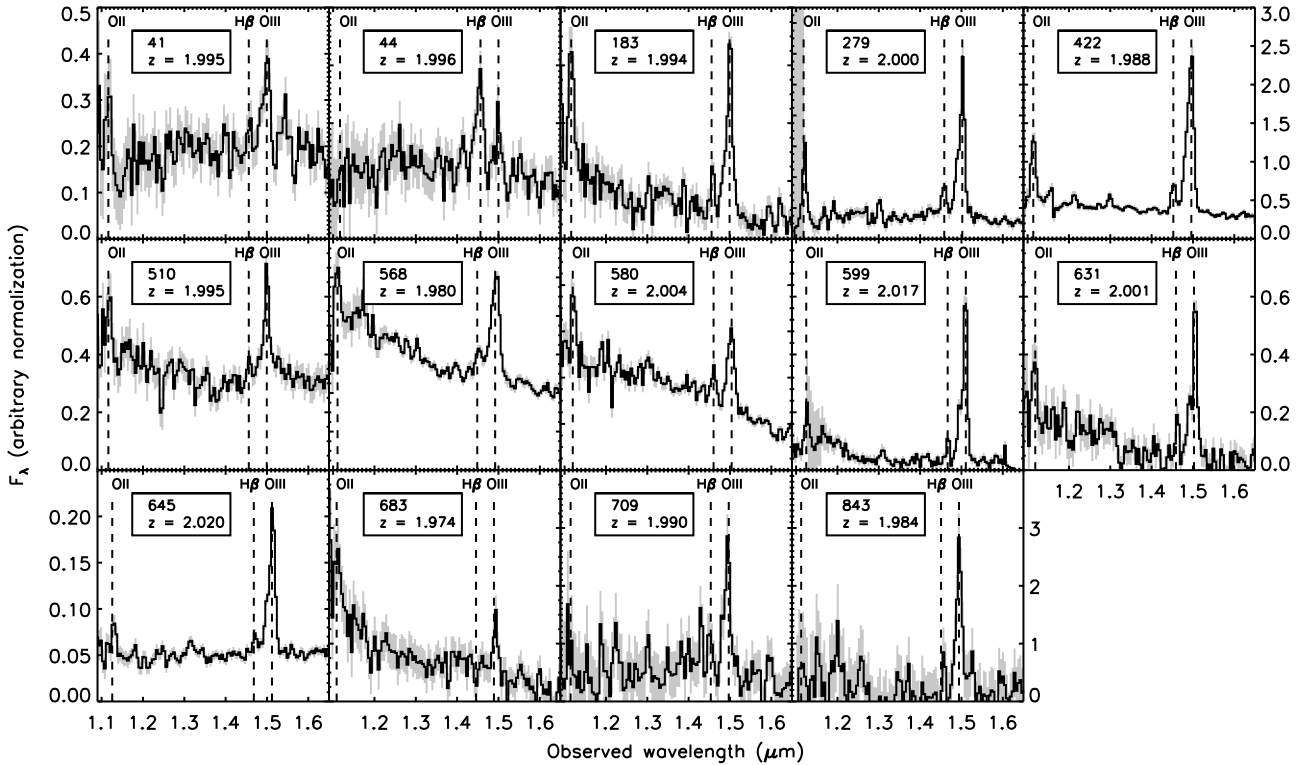}
\caption{\emph{HST}/WFC3 of emission-line members in Cl J1449+0856, with flux errors in gray. The position of prominent emission 
features is shown by dashed lines.}
\label{fig:linespec}
\end{figure*}

\section{\label{redshifts}Redshift determinations}

The usable range of the G141 is 1.1~$\mu$m--1.7~$\mu$m. At $z\sim2$ and considering the spectral resolution, observable features 
include [OII], [NeIII], H$\beta$ and [OIII]5007\AA~in emission, as well as the 4000\AA~break and less prominent Balmer lines, 
in the case of strongly star-forming or post-starburst galaxies. We have estimated redshifts for 140 targets in the field out of 
474 objects with $m_{140}<25.5$, with an average success rate of 30\%. The remaining objects, for which we could not recover a 
redshift were either irremediably contaminated (e.g., aligned, in the dispersion direction, with bright sources such as stars), 
had no significant spectral features in the G141 wavelength range (due to their redshift), or had too low S/N, either because 
of intrinsic faintness or low surface brightness.
The success rate is high at the center, as shown in Figure \ref{fig:success}, and declines toward the edge of the field. This is 
likely due to higher noise because of the lower coverage and also to the presence of relatively bright sources (low-redshift 
galaxies and stars) near the edges of the field.
Redshifts were estimated in two ways: by matching the position of peaks in the spectra with known emission lines and, for spectra 
without obvious emission features, by cross-correlation with stellar population templates.\\

\subsection{\label{lines}Line emitters}

Spectra were first automatically inspected for emission lines with S/N$>3$ and width that matched the effective resolution (i.e., the 
width of the source in the dispersion direction). Those that were close to zeroth orders were ignored, regardless of any differences 
between the observed flux and that predicted by the contamination model. A Gaussian fit was then done to the peaks and the centroids used 
as their ``true'' position. Here we used spectra extracted by the standard pipeline, as emission-line spectra do not necessarily require 
a good flux calibration for the identification and examination of a spectrum in all orientations allows for differentiation between lines 
arising from the source itself (which will be common to the different orientations) and from contamination. 
Two different quality flags were assigned to resulting redshifts: when two or more emission lines were found 
and their relative position matched, the redshift was deemed secure (``A'' quality) and the uncertainty on the redshift was determined by 
the scatter of line positions. Figure \ref{fig:linespec} shows the \emph{HST}/WFC3 emission-line spectra of star forming galaxies in 
Cl J1449+0856, counting close pairs, possibly clumpy objects, only once.
When only one peak could be reliably measured, and no other prominent emission lines were expected to fall in the grism range, the photometry 
was used to distinguish between possible solutions: first, a rough selection was done where the line was assumed to be [OIII]5007\AA~for 
sBzK-selected galaxies and H$\alpha$ otherwise. These single-line redshifts were then compared with the photometric redshifts 
\citep[see][]{Stra13}. In six cases, where the spectroscopic redshift was incompatible with the photometry, we chose the bright line that 
yielded the redshift most consistent with the photometric one (for example, [OII]3727\AA~instead of [OIII]5007\AA, especially when 
$z_{phot}\gtrsim2.5$).
Such redshifts were considered to be less secure (``B'' quality). In this case, the error on the redshift is proportional to the ratio 
of the line FWHM over the S/N.\\ 

Emission-line redshifts thus determined range from $z=0.3$ (from Paschen-series lines) to $z=3.1$ (from [OII]3727\AA). 
Our 3$\sigma$ line flux limit is $\sim5\times10^{-18}$~erg s$^{-1}$ cm$^{-2}$, without noticeable correlation of the S/N with either 
F140W magnitude or redshift. We find 46 H$\alpha$ emitters, in the range $z=0.7$--1.5, and 48 H$\beta$ and [OIII] emitters between $z=1.2$ 
and $z=2.1$, not counting cluster members. Figure \ref{fig:success} shows the cumulative distribution of these field [OIII] emitters as a 
function of line luminosity, as this latter population is perhaps more interesting in the context of future surveys such as Euclid 
\citep[e.g.,][]{Euc}, which will have a spectroscopic wavelength coverage similar to WFC3. Note that Figure \ref{fig:success} shows the 
measured, uncorrected counts and represents a lower limit: while all the sources are brighter than the photometric completeness limit, and 
we can thus be reasonably certain that we are not missing any candidates, correcting for the spectroscopic incompleteness, which in this 
case depends on contamination and thus the distribution of bright sources, is trickier. Ignoring cosmic variance and assuming a success 
rate similar to ours, we can estimate that there should be up to $\sim$600 (respectively $\sim$6000) such detectable line emitting galaxies 
per deg$^2$ in the Euclid wide (respectively deep) survey.\\

\begin{figure}
\centering
\includegraphics[width=0.49\textwidth]{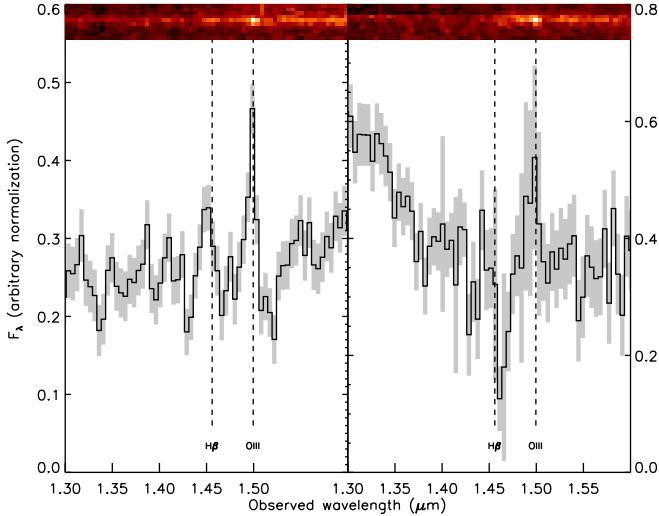}
\caption{Spectra of the two X-ray detected AGNs, showing the [OIII]5007\AA~emission. The spectrum of the right one is too contaminated at 
$<14000$\AA~to be corrected fully, but the [OIII] line appears in all three orientations.}
\label{fig:agn}
\end{figure}

\subsubsection{\label{agn}Active Galactic Nuclei}

Two of the galaxies near the cluster core are X-ray sources, one detected in the soft band \citep[as reported in][]{Gob11} and 
the other, tentatively, in the hard band. The spectra of these two sources have detectable [OIII]5007\AA~emission, as shown in Figure 
\ref{fig:agn}, and their redshifts were thus determined from these lines as described above. The position of the soft X-ray source is 
actually occupied by a close pair of compact objects, within the same \emph{Chandra} PSF, and we are not able to determine with confidence 
whether the emission stems from one of the sources or both. We hereafter attribute the redshift to the brightest one. 
The (tentative) hard X-ray detection is classified as an ETG on the basis of color and morphology and is included in the 
sample of \citet{Stra13}. However, we could not spectroscopically confirm its quiescent nature, as its spectrum is unusable below 
14000\AA, being too contaminated in all three orientations.\\

\subsubsection{\label{comp}Comparison with Ground-based Spectra}

\begin{figure}
\centering
\includegraphics[width=0.49\textwidth]{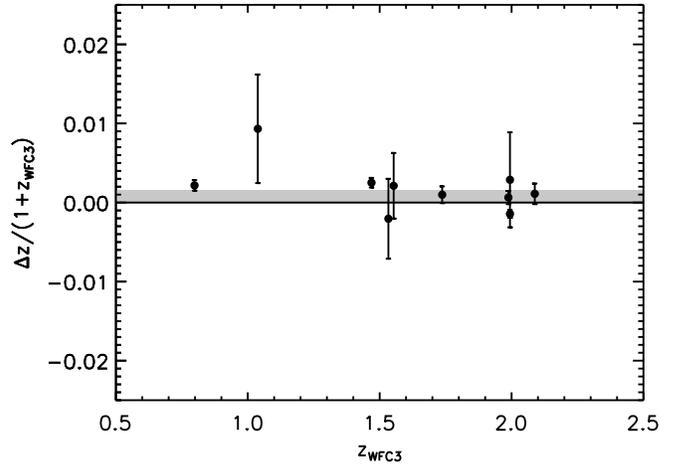}
\caption{Difference between \emph{HST}/WFC3 and VLT/FORS+VIMOS redshifts, with the average offset and scatter shown in gray.}
\label{fig:zcomp}
\end{figure}

There are 10 galaxies in the \emph{HST}/WFC3 field which have redshifts determined both from G141 and ground-based (VLT/FORS 
and VLT/VIMOS) spectra, which allowed us to check the reliability of our WFC3-derived redshifts. We indeed find them to be 
consistent, as shown in Figure \ref{fig:zcomp}, with a scatter of $\sigma_{{\rm WFC3}}=0.003$ corresponding to a fraction of a WFC3 
pixel.\\

\begin{figure}
\centering
\includegraphics[width=0.49\textwidth]{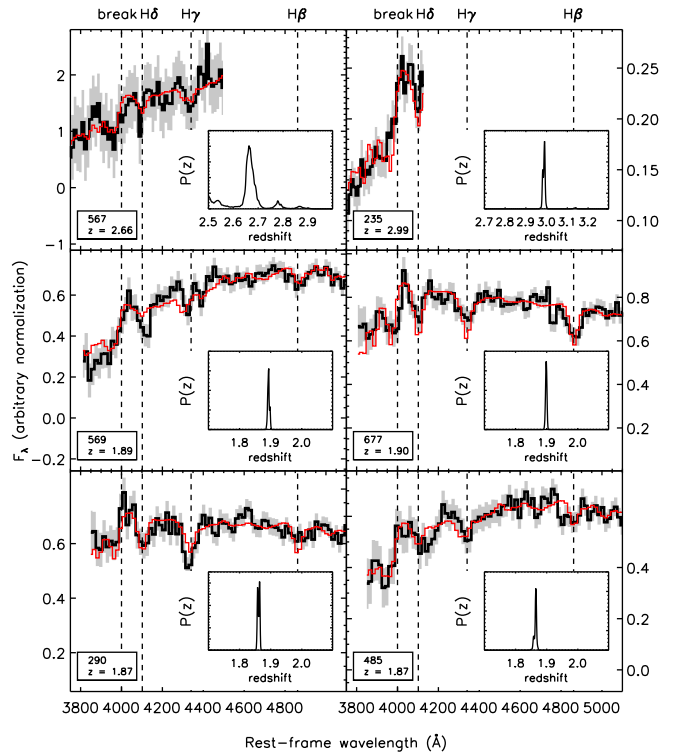}
\caption{Spectra of quiescent field galaxies, in the rest frame, with best-fit models in red and the position of detectable spectral 
features shown by dashed lines. The inserts show, for each galaxy, the $\chi^2$ probability distribution as a function of redshift.}
\label{fig:etgfield}
\end{figure}

\begin{figure*}
\centering
\includegraphics[width=0.98\textwidth]{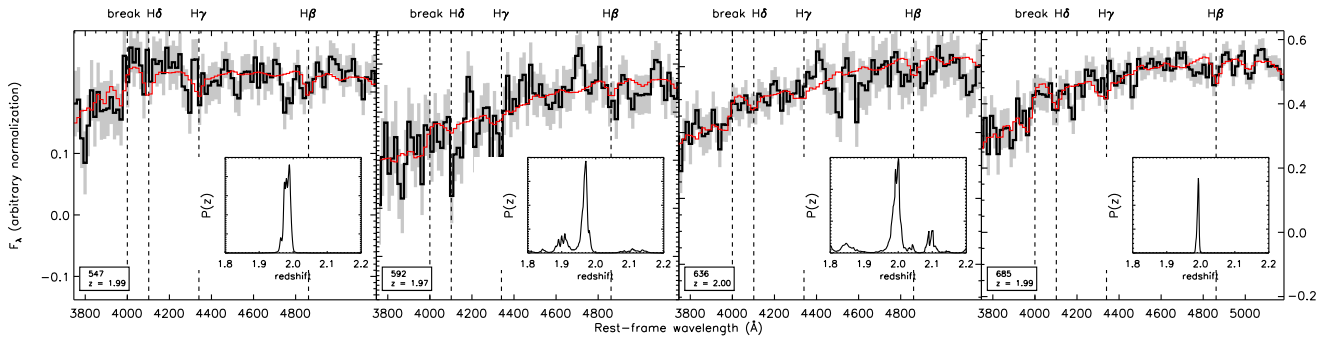}
\caption{Spectra of the four ``good'' non-AGN quiescent members of Cl J1449+0856, in the rest frame, with best-fit models in red and 
the position of detectable spectral features shown by dashed lines. As in Figure \ref{fig:etgfield}, the inserts show, for each galaxy, the 
$\chi^2$ probability distribution as a function of redshift.}
\label{fig:etgclu}
\end{figure*}

\subsection{\label{redgal}Early-type and Non Emission-line Galaxies}

The core of Cl J1449+0856 is dominated by a red galaxy population \citep{Gob11}.
From a purely photometric analysis, this results to be a mix of quiescent stellar populations and highly reddened systems, the spectra 
of which mostly lack emission lines with the exception of the active galactic nucleus (AGN) mentioned above in Section \ref{agn}. 
This initial selection, described in \citet{Stra13}, was based on a combination of rest-frame $U-V$ and $V-J$ colors \citep{Wuy07} and SED 
modeling. The quiescent or dusty nature of the spectroscopic members has been confirmed by the analysis described here, except for the AGNs 
described above in Section \ref{agn}.\\

For these red galaxies, the method described in Section \ref{ext} yielded the best results, while the output of the standard extraction was more 
difficult to interpret. 
As shown in Figure \ref{fig:grism}, there are five non-AGN quiescent galaxies in the 
core with usable spectra. For these, as well as the rest of red continuum objects, we estimated redshifts by fitting stellar 
population templates \citep{BC03}, rebinned to the grism resolution and broadened by the galaxy's profile in the case of extended 
sources. We assumed delayed, exponentially declining star formation histories and included line emission based on the models' star 
formation histories as well as dust extinction, similar to the procedure described in \citet{Gob12}. 
Each fit was then visually inspected and the best-fit model compared to the SED. In a few cases where there was a significant 
discrepancy between the model and the SED, or when the fit yielded significant secondary solutions (e.g., double-peaked probability 
distributions), a combined fit was done including the SED. In the former case, the combined fit produced a best-fit stellar population 
model more consistent with the observations but did not change the redshift very much, while in the latter it allowed us to discard 
possible redshift solutions.
The quality of these redshifts was assigned after this inspection: when the spectrum had evident, easy to interpret features and the 
uncertainty on the redshift measurement was small ($\sim0.02$), the fit was considered secure (``A'' quality). On the other hand, in 
cases where the SED was used, or when the redshift error was large (e.g., $\sim0.03$), the estimate was deemed less secure 
(``B'' quality).\\

We were thus able to obtain redshifts for 11 quiescent non-AGN galaxies, both in the cluster and in the field, from $z=1.86$ to 
$z=2.99$ \citep[the latter being described in][]{Gob12}, as well as a dusty system in the cluster core. 
Interestingly, $z\sim1.9$ interlopers were the easiest to confirm, as they had relatively high S/N spectra and well defined absorption 
features \citep[which could result from their small sizes; see][]{Stra13}, as shown in Figure \ref{fig:etgfield}. More importantly, we could 
spectroscopically confirm the redshift of the quiescent candidates in the cluster core, which had so far been unfeasible, and thus 
confirm without doubt the redshift of Cl J1449+0856. Their spectra are shown in Figure \ref{fig:etgclu}. 
The spectroscopically confirmed quiescent galaxies have morphologies that can be described with a $n>2$ Sersic profile and are thus 
classified as early-types in \citet{Stra13}, with one exception: a quiescent cluster member, which is part of the central interacting system 
\citep[referred to as a ``proto-BCG'' in][]{Gob11} and for which a reliable morphological fit could not be obtained.\\

In addition, two of the red galaxies in the cluster core have extended profiles which degrade the effective spectral resolution (one being 
an ETG with an extended halo and the other a dusty star-forming system with clear merger-like features, respectively; IDs 583 and 621 in 
Table \ref{tab:mem}). The fit on their spectra and SEDs 
independently and consistently produced single $z=2$ (respectively quiescent and dusty) solutions and we include them as cluster members. 
However, we consider their redshifts to be of low quality, due to the low S/N and resolution, and do not include the quiescent one in the 
spectral stack and the subsequent analysis (see below).\\

\begin{figure}
\centering
\includegraphics[width=0.49\textwidth]{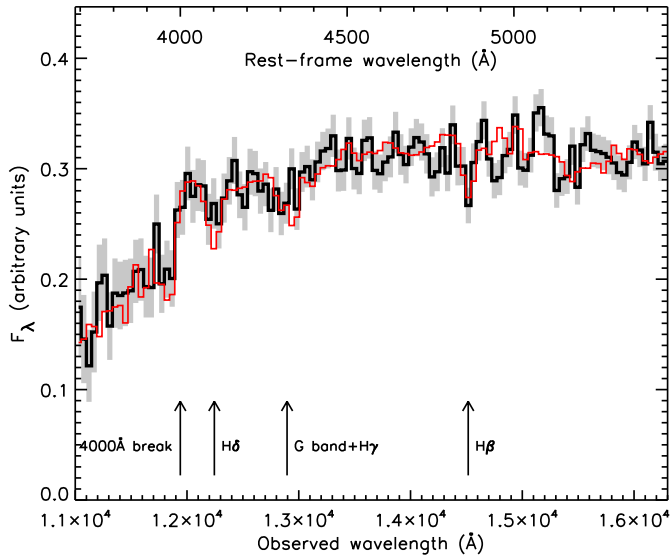}
\caption{Stacked spectrum of four quiescent galaxies in Cl J1449+0856 (in black), seen in Figure \ref{fig:etgclu}, with the best-fit model 
in red (1~Gyr at solar metallicity).}
\label{fig:etgstack}
\end{figure}

While spectral extraction was performed on all photometrically selected quiescent galaxies in the \emph{HST}/WFC3 field, with a photometric 
redshift consistent with the presence of the 4000\AA~break in the spectral range, no other object yielded a usable spectrum. 
Due to either too low S/N or too high contamination, the redshift of these objects could not be measured with 
any degree of confidence. We note that most of the spectroscopically confirmed quiescent galaxies, save for two compact cluster members, 
belong to the high-mass end of the galaxy population at $z\sim2$, with $M\gtrsim10^{11}$~$M_{\sun}$ \citep{Stra13}.
As a sanity check, we produced a high-quality stack from the spectra, in each orientation, of these four ``good'' quiescent cluster members. 
As shown in Figure \ref{fig:etgstack}, this stacked spectrum displays clear features that are less prominent in the individual spectra.\\

\begin{figure*}
\centering
\includegraphics[width=0.49\textwidth]{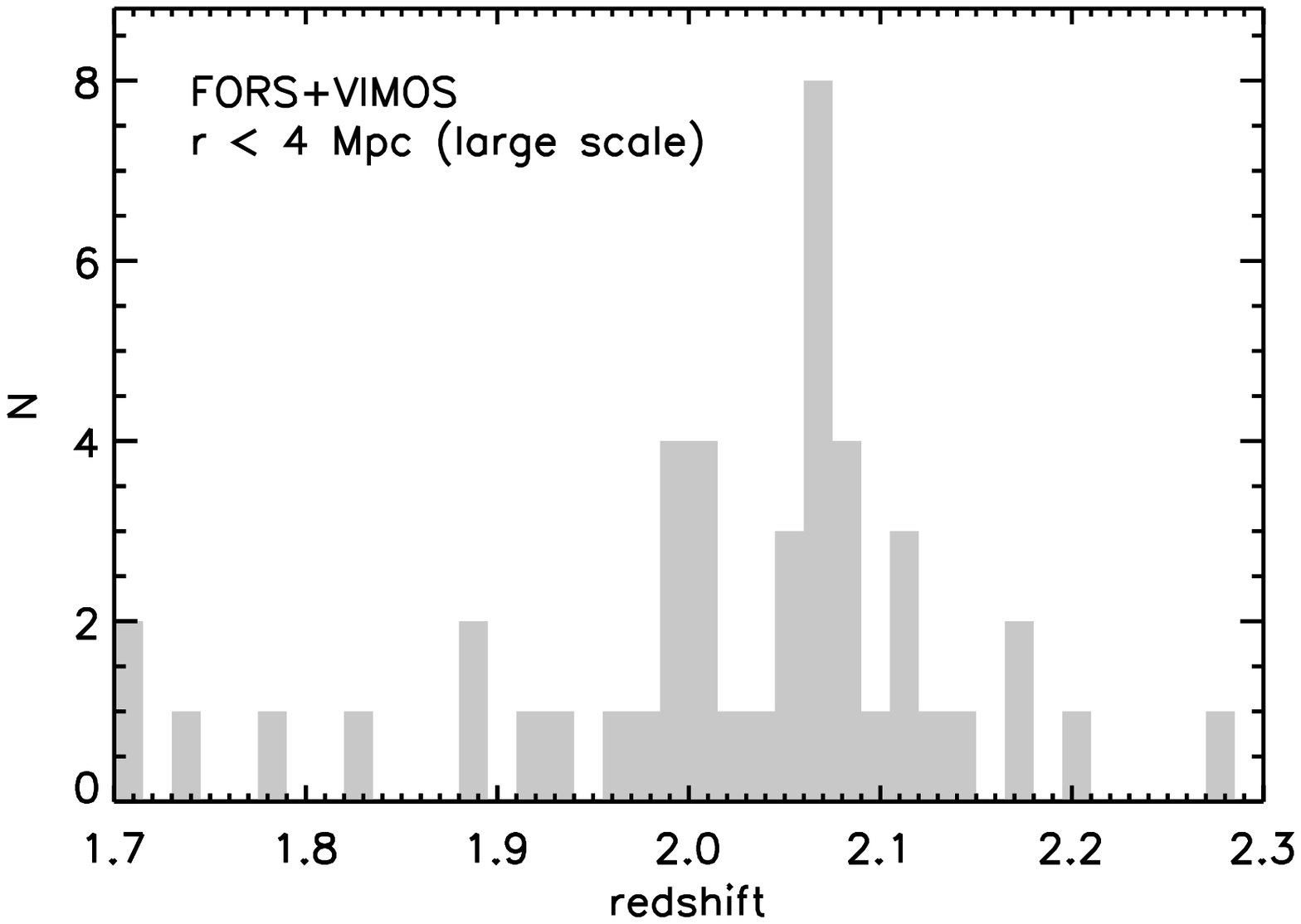}
\includegraphics[width=0.49\textwidth]{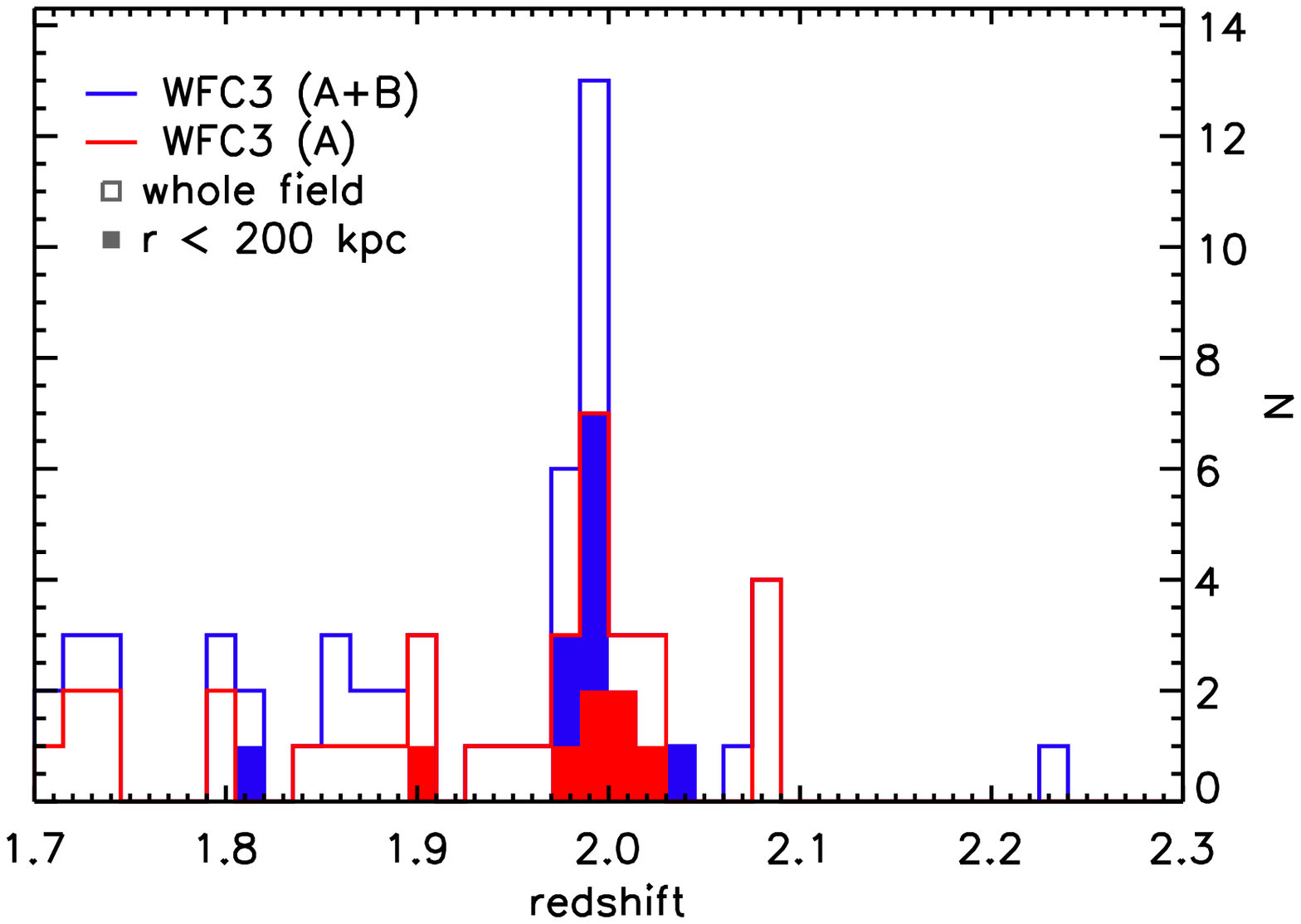}
\caption{Left: distribution of redshifts in the FORS and VIMOS masks previously published in \citet{Gob11}. 
Right: distribution of \emph{HST}/WFC3 G141 redshifts, where the $z=2.00$ peak is much more prominent. The blue distributions 
include all redshifts, while the red ones show only the better quality redshifts. Open histograms show the redshift distributions 
for the entire \emph{HST}/WFC3 field while the solid ones include only objects within 200~kpc of the cluster center.}
\label{fig:hz}
\end{figure*}

\section{\label{results}Results}

Having obtained the redshifts of quiescent galaxies in the cluster center, as well as active galaxies throughout the field, we can 
now securely confirm the redshift of Cl J1449+0856. However, the distribution of emission-line WFC3 redshifts peaks strongly at $z=2.00$, 
as shown in Figure \ref{fig:hz}, instead of $z=2.07$ as reported in \citet{Gob11} using VLT spectra of sBzK-selected galaxies outside 
the cluster core. 
Also spectroscopically confirmed quiescent galaxies in the cluster core strongly peak at $\langle z\rangle =1.99$. In particular, two 
components of the ``proto-BCG'' have consistent $z\sim2$ redshifts, although only one of them is secure (``A'' quality).
We thus conclude that Cl J1449+0856 is indeed a real structure, but a little closer than previously thought. The full distribution, 
including the redshifts from the previous, ground-based spectroscopy, has a biweight mean of $\langle z\rangle =1.995$ with a scatter 
of 0.013. We thus hereafter quote $z=2.00$ as the redshift of the cluster.
The scatter would imply a velocity dispersion of $\sim1300$~km s$^{-1}$, incompatible with the relatively low mass 
($\sim5\times10^{13}$~$M_{\sun}$) derived from the cluster X-ray emission and richness. Although, in principle, it may suggest the presence 
of substructure, we note that this scatter is close to the accuracy reached for red galaxies and ``B'' quality emission-line redshifts 
and that it is thus very likely dominated by redshift uncertainties.
Accordingly, given the relatively large uncertainty on the redshift of individual galaxies and resulting scatter around the $z=2$ 
peak, we consider as cluster member galaxies in the range $1.97<z<2.02$.\\

\begin{figure*}
\centering
\includegraphics[width=0.98\textwidth]{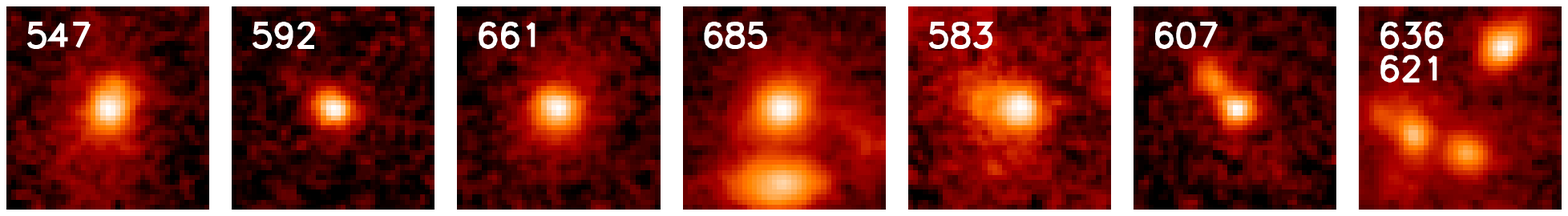}
\includegraphics[width=0.98\textwidth]{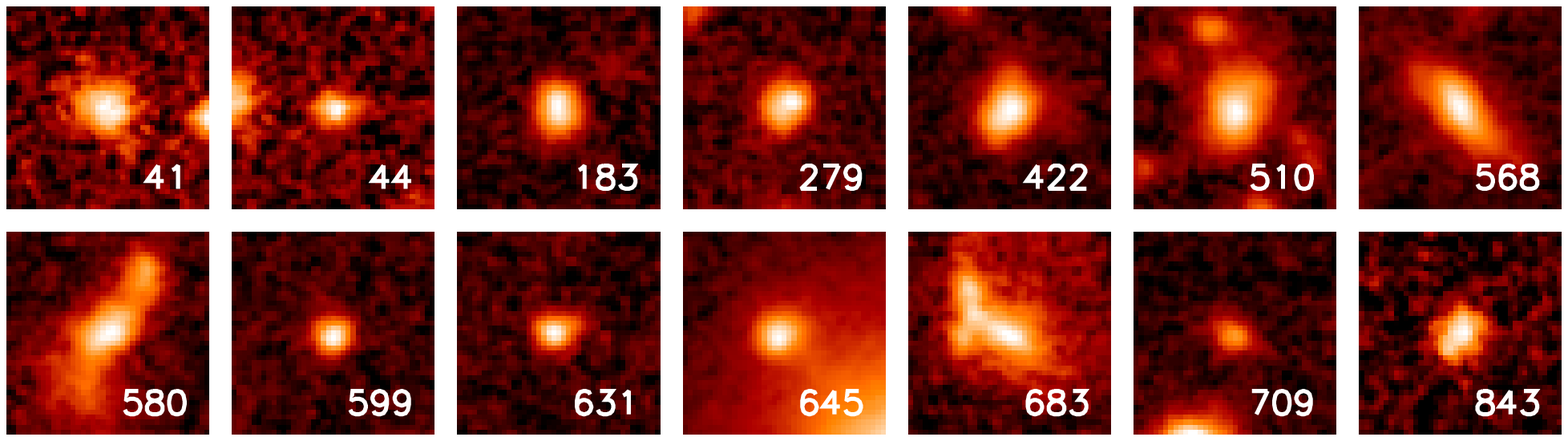}
\caption{WFC3 F140W cutouts of cluster members, with the ID in white. 
The top row shows the red spectroscopic members, also shown in \citet{Stra13} and described in Table \ref{tab:mem}, while emission-line members 
are shown in the bottom two rows. Close pairs with matching redshifts are shown only once. The cutout size is 2.\arcsec4~or $\sim21$~kpc at 
$z=2$.}
\label{fig:cutouts}
\end{figure*}

We have spectroscopically confirmed 22 objects at $z\sim2$, including 5 quiescent galaxies, 2 emission line AGN and 16 star-forming 
galaxies (1 dusty and 15 line-emitting), counting close pairs as a single source.
Images of the spectroscopic members, in the \emph{HST}/WFC3 F140W band, are shown in Figure \ref{fig:cutouts}.
In addition, there are 5 more galaxies in the range $1.97<z<2.02$ not in the WFC3 dataset but with ground based redshifts, 
for a total of 27 spectroscopic members, as listed in Table \ref{tab:mem}.\\

\begin{figure*}
\centering
\includegraphics[width=0.98\textwidth]{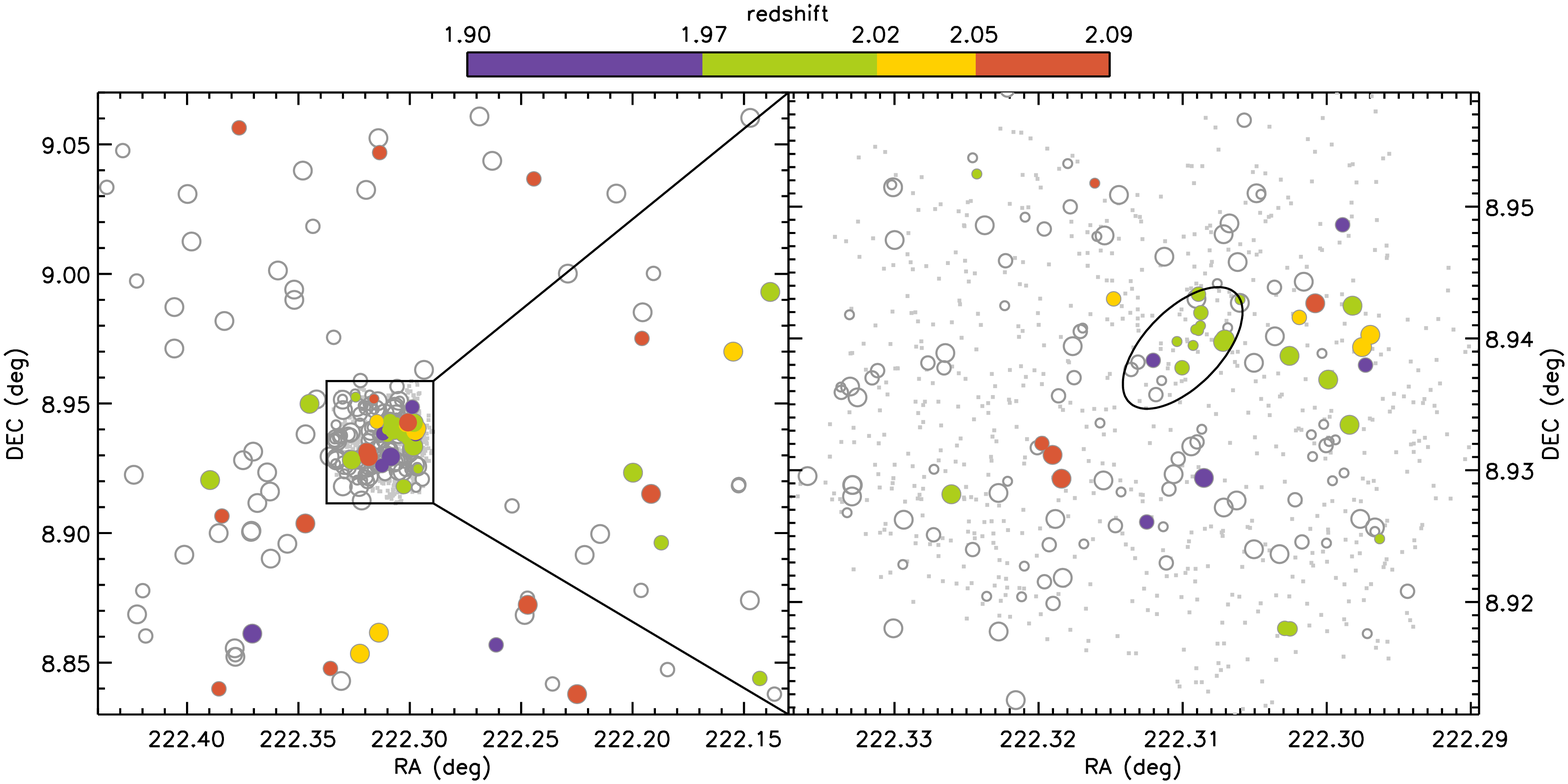}
\caption{Positions of galaxies with redshifts between $z=1.9$ and 2.1 showing, respectively, the distribution of 
spectroscopic members of Cl J1449+0856 (green) and the $z=2.07$ structure (orange). Left: galaxies in both the \emph{HST}/WFC3 
and FORS+VIMOS fields with G141 and ground-based redshifts. The black rectangle corresponds to the area shown on the right.
Right: galaxies in the \emph{HST}/WFC3 field with G141 spectra. Dots show the position of sources without redshifts and empty 
circles that of sources with redshifts outside the range. The size of the symbols is proportional to redshift quality and the S/N of 
the spectra.}
\label{fig:map}
\end{figure*}

As shown in Figure \ref{fig:map}, the spatial distribution of galaxies in the $z=2.00$ peak is quite different than that of those 
in the $z=2.07$ one. They form a more concentrated structure, perhaps also slightly elongated in nature, although this could be an 
artifact due to the number of bright sources (and thus potentially catastrophic contaminants) being higher on one side of the field. 
Galaxies at $z=2.07$ lie in a sparser, more extended and possibly ``sheet''-like structure in the background.\\

The previous misidentification, and redshift distributions seen in Figure \ref{fig:hz}, can then be easily explained by the 
different observing capabilities of the two types of instruments used here and in \citet{Gob11}: 
the field of view of FORS and VIMOS is 10--20 times wider than the extension of Cl J1449+0856, which is a few hundred 
kiloparsecs across and about one arcminute on the sky. Moreover, as slits cannot be placed closer than several arcseconds, these 
instruments can sample only very few galaxies in the cluster core and, instead, are more effective at revealing the presence of 
redshift spikes extending over many arcminutes. On the other hand, the WFC3 is much more effective in handling dense fields, such 
as a cluster core.\\

It is certainly quite common to see several overdensities in redshift space along the same line of sight, though in this 
case the cluster at $z=2.00$ and the {\it sheet} at $z=2.07$ appear to be very close to each other. The $z=2.07$ structure appears 
to be much less dense than Cl J1449+0856 and to lack a red galaxy population. On the other hand, cluster and sheet are not 
gravitationally bound, being separated by $\sim100$~Mpc (comoving). As this implies an infall time larger than the age of the Universe, 
the two structures, while part of the same portion of ``cosmic web'', are unlikely to coalesce by $z=0$.\\

Furthermore, we note that the slight redshift revision does not change much the mass of Cl J1449+0856: the intrinsic X-ray 
luminosity becomes $\sim10$\% dimmer 
\citep[$L_X(0.1-2.4$~keV$)=6.4\pm{1.8}\times10^{43}$~erg~s$^{-1}$, assuming the same scaling relations as in][]{Gob11}, 
corresponding to a mass $\sim5$\% lower, of $M_{200}=5.3\pm{0.9}\times10^{13}$~$M_{\odot}$, i.e., within the uncertainty of the first 
estimate. Likewise, the inferred  radius does not change appreciably compared to our previous estimate as $r_{200}=0.4$~Mpc.\\

\begin{figure}
\centering
\includegraphics[width=0.49\textwidth]{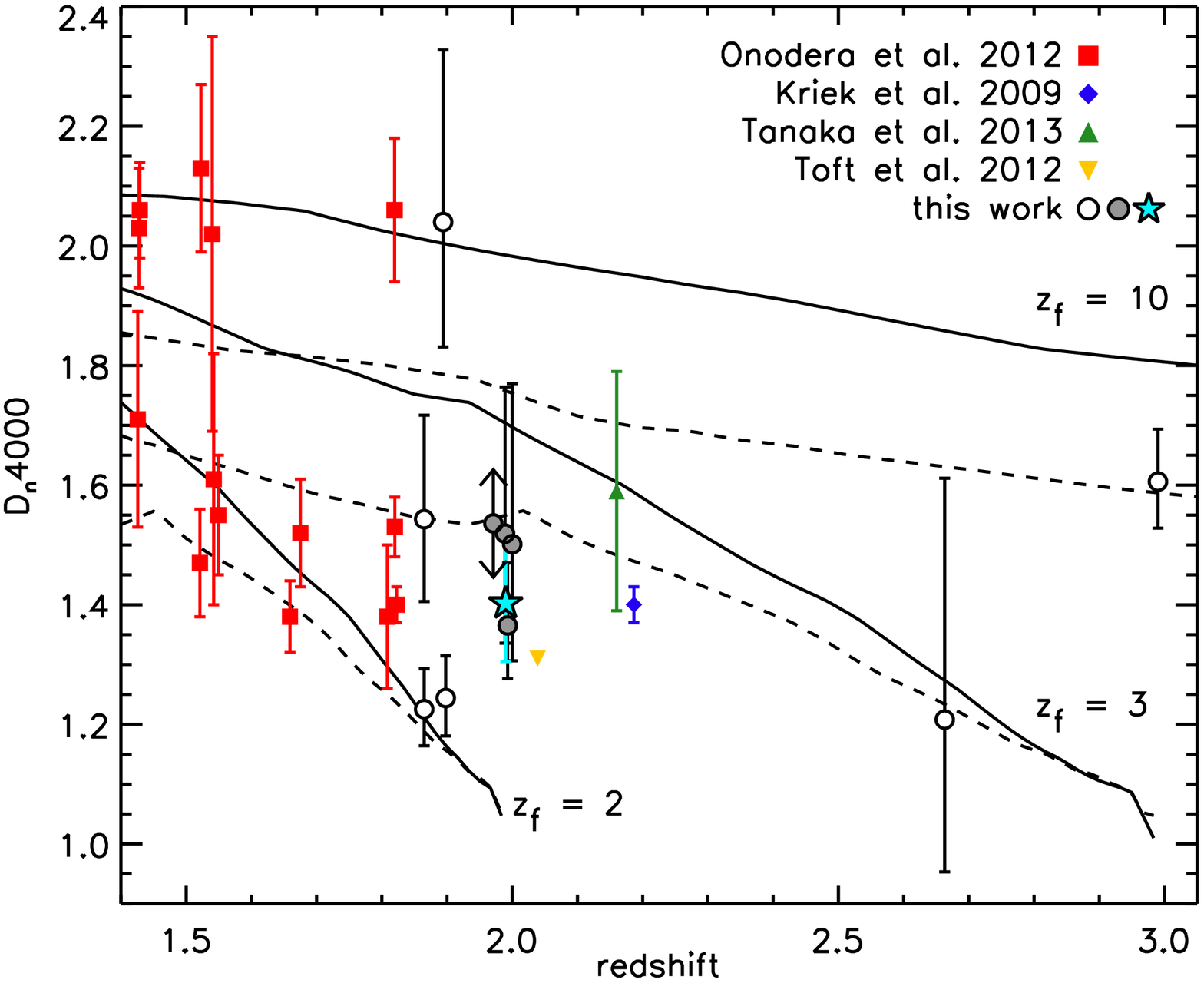}
\caption{$D_{\textrm{n}}$4000 index as a function of redshift of field and cluster (respectively open and 
filled black circles, with the cluster stack as a light blue star) quiescent galaxies, compared to other high redshift field ETG samples 
\citep{Tof12,Ono12,Kri09} as well as \citet{Tan13} proto-cluster stack. Error bars have been replaced by arrows for objects with S/N$<3$.
The effects of low resolution on the $D_{\textrm{n}}$4000 were corrected using the best-fit models to the G141 spectra. 
As a visual aid, the dashed and solid lines show the $D_{\textrm{n}}$4000 value, as a function of redshift, of \citet{BC03} single 
stellar population templates of solar and twice solar metallicity formed at $z_f=2,3$~and 10 (i.e., maximally old).}
\label{fig:d4k}
\end{figure}

Finally, we compare age indicators for the quiescent cluster members, individual and stacked, with the field sample. We find that the 
stacked spectrum is consistent with a stellar population of star-formation-weighted age $\sim1$~Gyr (i.e., a formation redshift of $z_f\sim3$), 
assuming solar metallicity, with little extinction ($E(B-V)\lesssim0.1$). This is in agreement with the strength of its 4000\AA~break
\citep[$D_{\textrm{n}}4000=1.4\pm0.08$ as quantified by the $D_{\textrm{n}}$4000 index][]{Balo99} and that of the individual quiescent cluster 
galaxies. We find that these values are consistent with those of field ETGs at $z\sim1.9$ in our sample and others \citep{Tof12,Ono12,Kri09}, 
as shown in Figure \ref{fig:d4k}. In apparent contrast with some observations of lower redshift, more evolved clusters 
\citep[e.g.,][but see also \cite{Muz12}]{Gob08,Ros09,Ret11}, there appears to be no detectable age difference (barring metallicity 
effects) between cluster and field quiescent galaxies. If true, this would suggest a similarity of stellar populations in already quenched 
galaxies in different environments at $z\sim2$.

\begin{deluxetable*}{cccccccccc}
\tablecaption{Spectroscopic Cluster Members\label{tab:mem}}
\tablehead{\colhead{ID} & \colhead{RA} & \colhead{Decl.} & \colhead{$m_{140}$} & \colhead{Redshift} & \colhead{Error} & \colhead{Quality} & 
\colhead{Instrument} & \colhead{Method\tablenotemark{a}} & \colhead {Object Type\tablenotemark{b}}\\
\colhead{} & \colhead{(deg)} & \colhead{(deg)} & \colhead{} & \colhead{} & \colhead{} & \colhead{} & \colhead{} & \colhead{} & 
\colhead{}}
\startdata
547&222.3100399&8.9377762&23.42&1.98&0.02&A&WFC3&fit&Quiescent\\
583&222.3092734&8.9394681&22.95&2.00&0.03&B&WFC3&fit&Quiescent,low S/N\\
592&222.3106699&8.9390994&24.73&1.97&0.02&B&WFC3&fit&Quiescent\\
636&222.3087650&8.9409647&23.60&1.99&0.03&B&WFC3&fit&Quiescent\\
685&222.3060186&8.9429459&22.60&1.98&0.02&A&WFC3&fit&Quiescent\\
661&222.3087325&8.9419476&23.44&2.002&0.001&B&WFC3&line&Quiescent,X-ray AGN\\
607&222.3103930&8.9397659&24.30&1.992&0.005&B&WFC3&line&X-ray AGN\\
621&222.3090911&8.9406764&22.67&1.99&0.02&B&WFC3&line&SF,dusty,low S/N\\
41&222.3029141&8.9180000&24.01&1.995&0.002&A&WFC3&line&SF\\
44&222.3025456&8.9179633&24.79&1.996&0.001&A&WFC3&line&SF\\
183&222.2963297&8.9247956&23.44&1.994&0.020&B&WFC3&line&SF\\
279&222.3260510&8.9281756&23.88&2.000&0.001&A&WFC3&line&SF\\
422&222.2984180&8.9334464&22.76&1.988&0.002&A&WFC3&line&SF\\
510&222.2998987&8.9368734&22.77&1.995&0.001&A&WFC3&line&SF\\
568&222.3025838&8.9386765&22.46&1.980&0.004&A&WFC3&line&SF\\
580&222.3071947&8.9397384&22.75&2.004&0.001&A&WFC3&line&SF\\
599&222.2975461&8.9393429&24.52&2.017&0.003&A&WFC3&line&SF\\
631&222.2969763&8.9402737&24.73&2.001&0.009&A&WFC3&line&SF\\
645&222.3019033&8.9415953&22.84&2.020&0.010&A&WFC3&line&SF\\
683&222.2982125&8.9424841&24.01&1.974&0.030&B&WFC3&line&SF\\
709&222.3089050&8.9433599&24.99&1.990&0.009&A&WFC3&line&SF\\
843&222.3242837&8.9524799&24.43&1.984&0.001&A&WFC3&line&SF\\
22111&222.3450108&8.9498639&...&2.004&0.001&A&FORS2&fit&SF\\
1127&222.3224030&8.8534632&...&2.014&0.001&A&VIMOS&fit&SF\\
141.2&222.1997986&8.9233017&...&2.004&0.001&A&VIMOS&fit&SF\\
1411.3&222.1871033&8.8962622&...&1.991&0.002&B&VIMOS&fit&SF\\
1438&222.1428833&8.8439026&...&1.970&0.002&B&VIMOS&fit&SF\\
\enddata
\tablenotetext{a}{Spectrum fit or line identification}
\tablenotetext{b}{Quiescent, AGN, or star-forming (SF).}
\end{deluxetable*}

\section{\label{conc}Summary}

We have used deep grism observations with \emph{HST}/WFC3 to study the distribution of galaxies in a $\sim6.4$~arcmin$^2$ field 
centered on the distant cluster Cl J1449+0856. We have obtained redshifts for 140 out of 474 sources in the \emph{HST}/WFC3 field with 
extracted spectra, down to a F140W magnitude of 25.5, with a success rate of $\sim50$\% at $m_{140}\leq24$ in the uniform coverage area.
While these WFC3 observations are unusually deep, the crowding of the field and faintness of our targets presented a technical challenge. 
To recover usable spectra in spite of strong contamination, in some cases we had to use a more sophisticated extraction technique than 
currently implemented in the standard software. This allowed us to achieve the following main results:\\

$\bullet$ We have securely measured the redshift of Cl J1449+0856 by spectroscopically identifying many cluster members, including 
several ETGs in its core, which had so far proven to be unfeasible with ground-based observations. We have found that 
the redshift of $z=2.07$ previously published for this cluster actually stems from a chance alignment with a more diffuse, sheet-like structure 
in the background. The WFC3 observations, which cover the cluster up to its putative virial radius, instead reveal that the galaxies in 
the core, including the quiescent galaxies, strongly peak at $\langle z\rangle=2.00$. 
These galaxies had anyway been assigned to the cluster in our previous study based on photometric redshifts, whose uncertainty is 
larger than the separation of the two peaks. The cluster Cl J1449+0856 and the $z=2.07$ overdensity may be part of a same large-scale 
structure but, being $\sim100$~comoving Mpc apart, do not appear to be gravitationally bound to each other.\\
$\bullet$ The grism spectroscopy confirms the quiescent nature of the photometrically selected passive candidates in the cluster.
We have constructed a high S/N average spectrum of the quiescent cluster members which indicates an age of $\sim1$~Gyr for the bulk of 
stellar populations in these galaxies, in agreement with the 4000\AA~break strength of individual galaxies. Under the assumption of 
similar metallicity, this is consistent with the age inferred for the quiescent field galaxies at $z\sim2$.\\
$\bullet$ We have assembled the first sample of spectroscopically confirmed quiescent galaxies in a high density environment at 
$z=2$. We have used these objects together with ETGs in the field around the cluster to constrain the effect of environment on galaxy 
properties at $z\sim2$ \citep{Stra13}.\\

These results illustrate the importance of a thorough spectroscopic follow-up of the quiescent galaxy population in high-redshift 
clusters and cluster candidates. They also demonstrate the unique capabilities of the WFC3 instrument, especially for continuum science.
This is particularly relevant in light of the success of photometric cluster searches which are now effectively approaching the 
$z\sim2$ range. Yet, the analysis of slitless spectroscopic data becomes more challenging the deeper the observations and more 
crowded the field, hence an optimized reduction strategy is indispensable for maximizing the scientific return of \emph{HST} observations.\\

\acknowledgements
We thank P. Oesch for his constructive comments. R.G., V.S. and E.D. were supported by grants ERC-StG UPGAL 240039 and ANR-08-JCJC-0008. N.A. 
is supported by a Grant-in-Aid for Science Research (No. 23224005) by the Japanese Ministry of Education, Culture, Sports, Science and 
Technology. A.R. was supported through the INAF grant ``PRIN 2010". AC acknowledges the grants ASI n.I/023/12/0 ``Attivit\`{a} relative alla 
fase B2/Cper la missione Euclid'' and MIUR PRIN 2010-2011 ``The dark Universe and the cosmic evolution of baryons: from current surveys to 
Euclid''. This work is based on data collected under program GO-11648 with the NASA/ESA \emph{Hubble Space Telescope}, which is operated by 
the Association of Universities for Research in Astronomy, Inc., under NASA contract NAS 5-26555.

\end{document}